\documentclass[a4paper,10pt]{article}
\usepackage{graphicx}
\usepackage{comment}
\usepackage{amssymb}
\usepackage{amsmath}
\usepackage{nicefrac}
\usepackage{graphicx}
\usepackage{dcolumn}
\usepackage{bm}
\usepackage{comment}
\usepackage{multirow}
\usepackage{cite}
\usepackage{xcolor}
\usepackage{url}

\begin{document}

\newcommand{\bin}[2]{\left(\begin{array}{c}\!#1\!\\\!#2\!\end{array}\right)}
\newcommand{\threej}[6]{\left(\begin{array}{ccc}#1 & #2 & #3 \\ #4 & #5 & #6 \end{array}\right)}
\newcommand{\sixj}[6]{\left\{\begin{array}{ccc}#1 & #2 & #3 \\ #4 & #5 & #6 \end{array}\right\}}
\newcommand{\regge}[9]{\left[\begin{array}{ccc}#1 & #2 & #3 \\ #4 & #5 & #6 \\ #7 & #8 & #9 \end{array}\right]}
\newcommand{\La}[6]{\left[\begin{array}{ccc}#1 & #2 & #3 \\ #4 & #5 & #6 \end{array}\right]}
\newcommand{\hj}{\hat{J}}
\newcommand{\hux}{\hat{J}_{1x}}
\newcommand{\hdx}{\hat{J}_{2x}}
\newcommand{\huy}{\hat{J}_{1y}}
\newcommand{\hdy}{\hat{J}_{2y}}
\newcommand{\huz}{\hat{J}_{1z}}
\newcommand{\hdz}{\hat{J}_{2z}}
\newcommand{\hup}{\hat{J}_1^+}
\newcommand{\hum}{\hat{J}_1^-}
\newcommand{\hdp}{\hat{J}_2^+}
\newcommand{\hdm}{\hat{J}_2^-}

\huge

\begin{center}
Electron-impact ionization of Si IV-VIII relevant for inertial confinement fusion
\end{center}

\vspace{0.5cm}

\large

\begin{center}
Jean-Christophe Pain$^{a,b,}$\footnote{jean-christophe.pain@cea.fr} and Djamel Benredjem$^{c}$
\end{center}

\normalsize

\begin{center}
\it $^a$CEA, DAM, DIF, F-91297 Arpajon, France\\
\it $^b$Universit\'e Paris-Saclay, CEA, Laboratoire Mati\`ere en Conditions Extr\^emes,\\
\it 91680 Bruy\`eres-le-Ch\^atel, France\\
\it $^c$Laboratoire Aim\'e Cotton, Universit\'e Paris-Saclay, Orsay, France\\
\end{center}

\vspace{0.5cm}

\begin{abstract}
In this work, we investigate the ionization of silicon by electron impacts in hot plasmas. Our calculations of the cross sections and rates rely on the Coulomb-Born-Exchange, Binary-Encounter-Dipole and Distorted-Wave methods implemented in the Flexible Atomic Code (FAC), and are compared with measurements and other theoretical values. We use a semi-empirical formula for the cross section, which involves a small set of adjustable parameters. Configuration interaction is taken into account and is shown to affect the cross section at low energy, in particular for Si$^{3+}$. The rate coefficient is then expressed in terms of these parameters and is represented in a large temperature interval, up to 10$^8$ K. As expected, the agreement with measurements improves for increasing ion charges, confirming the applicability of our approach to hot plasma studies such as inertial-confinement fusion, and its reliability.
\end{abstract}

\section{Introduction}\label{sec1}

In order to achieve ignition in a conventional inertial-confinement-fusion (ICF) implosion, a spherical shell filled with a frozen deuterium-tritium (DT) layer is driven inward either by X-rays emitted inside a Hohlraum in which laser beams are focused (indirect drive) or by laser irradiation directly impinging on the target (direct drive) \cite{Atzeni2004}. The shell is accelerated by the ablation pressure resulting from these radiation sources, converting the driver energy into kinetic energy of the shell. Recently, major progress has been made in the understanding of ICF, thanks to research in high-energy-density plasmas and to well-diagnosed experiments at the National Ignition Facility (NIF). In particular, the ability to ignite a capsule using a MJ-class laser facility was demonstrated on the NIF in August 2021 \cite{Abu2024}. Silicon is used in ICF capsules to reduce fuel preheat and laser–plasma instabilities \cite{Goncharov2014}. It enters as a component of the ablator, and is often in non local thermodynamic equilibrium (NLTE). The modeling of NLTE plasmas, such as those encountered in inertial confinement fusion, requires to solve the steady state population balance equations, including radiative and collisional processes \cite{Ralchenko2016}. More precisely, the most important processes are electron-impact excitation and de-excitation, electron-impact ionization and three-body recombination, radiative recombination, radiative de-excitation (or spontaneous emission), auto-ionization and resonant capture. In the presence of a radiation field, four additional processes must be taken into account: photo-excitation, stimulated emission, photo-ionization and stimulated recombination. The electron-impact ionization cross section, which is the subject of the present work, depends on the energy of the incident colliding electron. It must be averaged over the Maxwellian distribution of free electrons to yield the rate coefficient, which enters the collisional matrix, along with the rates of the other processes.

At first sight, electron-impact ionization (EII) might seem to be a relatively simple collision process. An incoming electron strikes an atomic target and one or more electrons are ejected. With the advent of quantum mechanics, approximations of increasing relevance were developed but experiments have revealed that the ionization of many targets remains very complex \cite{Dolder1994}. In 1982, Crandall \emph{et al.} reported measured EII cross sections for single ionization of Mg$^+$, Al$^{2+}$, and Si$^{3+}$ \cite{Crandall1982}. The authors employed crossed beams of electrons and ions to study the absolute cross sections as a function of collision energy. At that time, the cross sections for Mg$^{+}$ and Al$^{2+}$ near threshold were found to be roughly 70\% of the predicted direct ionization cross sections, while the agreement was reasonable for Si$^{3+}$. Contributions to the total cross section by indirect processes, principally inner-shell-excitation autoionization, have been specifically identified in each case and compared with theoretical results. These comparisons demonstrated specific failures of the predictions which rely on addition of excitation cross sections to the direct ionization one. In 1998, Badnell \emph{et al.} calculated the EII cross sections using R-matrix, time-independent and time-dependent close-coupling methods \cite{Badnell1998}. The results of all three methods yielded ionization cross sections lying substantially above the experimental crossed-beams measurements of Crandall \emph{et al.}. More recently, Jonauskas studied single ionization by electron impact in the Si atom by performing level-to-level calculations \cite{Jonauskas2020,Jonauskas2020a}. He investigated both direct and indirect processes of the ionization from all levels of the ground configuration, and found that cross sections are strongly dependent on the initial level for which the ionization is considered. It turns out that the cross sections of the indirect process differ by more than a factor of two for the lowest and highest levels of the ground configuration. In addition, modeling shows that nearly 70 \% of the atoms in the beam occupy the levels of the $^3$P term. 

In this work, we use the Flexible Atomic Code (FAC) \cite{Gu2008,FAC} for the computation of the EII cross section. The detailed level-to-level electron-impact ionization cross sections can be calculated by the relativistic Distorted-Wave (DW, see for instance Ref. \cite{Zhang1990,Fontes1993}), by the Coulomb-Born-Exchange (CB) \cite{Takagishi1985}, and by the Binary-Encounter-Dipole (BED) methods \cite{Vriens1966,Kim1994}.
The main features of the different methods considered in this work and the principles of the considered experiments are briefly recalled in section \ref{sec2}. All the characteristics of the calculations are provided in Ref. \cite{Gu2008}. The cross sections for ions Si$^{3+}$ to Si$^{7+}$ are investigated in section \ref{sec3}. Calculated and measured rates obtained using our recently published fitting procedure \cite{Benredjem2024} are discussed in section \ref{sec3} and compared to experimental values inferred by Zeijlmans van Emmichoven \emph{et al.} \cite{Zeijlmans1993}. A particular attention is paid to their variation with temperature.

\section{Cross sections of Si$^{3+}$, Si$^{4+}$, Si$^{5+}$ and Si$^{6+}$}\label{sec2}

The lowest ion charge for which reliable measurements are available in the literature is Si$^{3+}$. In the following, we compare our calculations with measurements of Crandall \emph{et al.} \cite{Crandall1982}. In the experiment, the ions are provided by the ORNL-PIG multicharged ion source and crossed-beam apparatus. A set of measured parameters among which the electron and ion currents provide the \textit{measured} cross section as can be seen in Eq. (1) of Ref. \cite{Crandall1982}.

The ionization potentials of ions Si$^{+}$ to Si$^{7+}$ are given in Table \ref{ionipot}. We can see that the values calculated by the FAC code are very close to those in the NIST (National Institute of Standards and Technology) database \cite{nist} for the ions Si$^{3+}$ to Si$^{7+}$, but very different for the lowest charges (Si$^+$ and Si$^{2+}$).

\begin{table}[!ht]
\begin{center}
\begin{tabular}{lrrr}\hline
Ion & FAC & NIST & Relat. diff.\\\hline\hline
Si$^{+}$  & 0.18 & 16.35&195.644\\
Si$^{2+}$ & 19.40 & 33.49&53.280\\
Si$^{3+}$ & 46.29 & 45.14&2.516\\
Si$^{4+}$ & 164.82 & 166.77&1.176\\
Si$^{5+}$ & 203.56 & 205.28&0.841\\
Si$^{6+}$ & 245.12 & 246.57&0.590\\
Si$^{7+}$ & 302.23 & 303.59&0.449\\\hline
\end{tabular}
\end{center}
\caption{Ionization potentials of silicon ions (in eV), given by the FAC code or collected from the NIST database. The fourth column contains the relative difference: $|$FAC-NIST$|/$((FAC+NIST)/2) in $\%$.}\label{ionipot}
\end{table}

In the DW method, the ionization radial integrals are calculated by summing up partial-wave contributions. This is a time consuming process because two continuum electrons are now involved. In the CB method, the radial integrals are simply looked up in a table. Consequently, the CB method is the fastest one. In the BED method, the radial integrals are calculated by using the bound-free differential oscillator strengths, which can be calculated much faster than the ionization radial integrals involved in the DW method.

In conjunction with the work of Crandall \emph{et al.}, CB (including excitation-autoionization) \cite{Moores1980} and DW \cite{Younger1980,Kim1994} calculations were also performed. In the CB method, the radial integrals were obtained by looking up a table of CB results from references \cite{Golden1977,Golden1980}, which is very fast. At that time, the experimental values were lower than CB and DW calculations. The agreement between theory and experiment was poor.

\begin{figure}
\centering
\includegraphics[scale=.45]{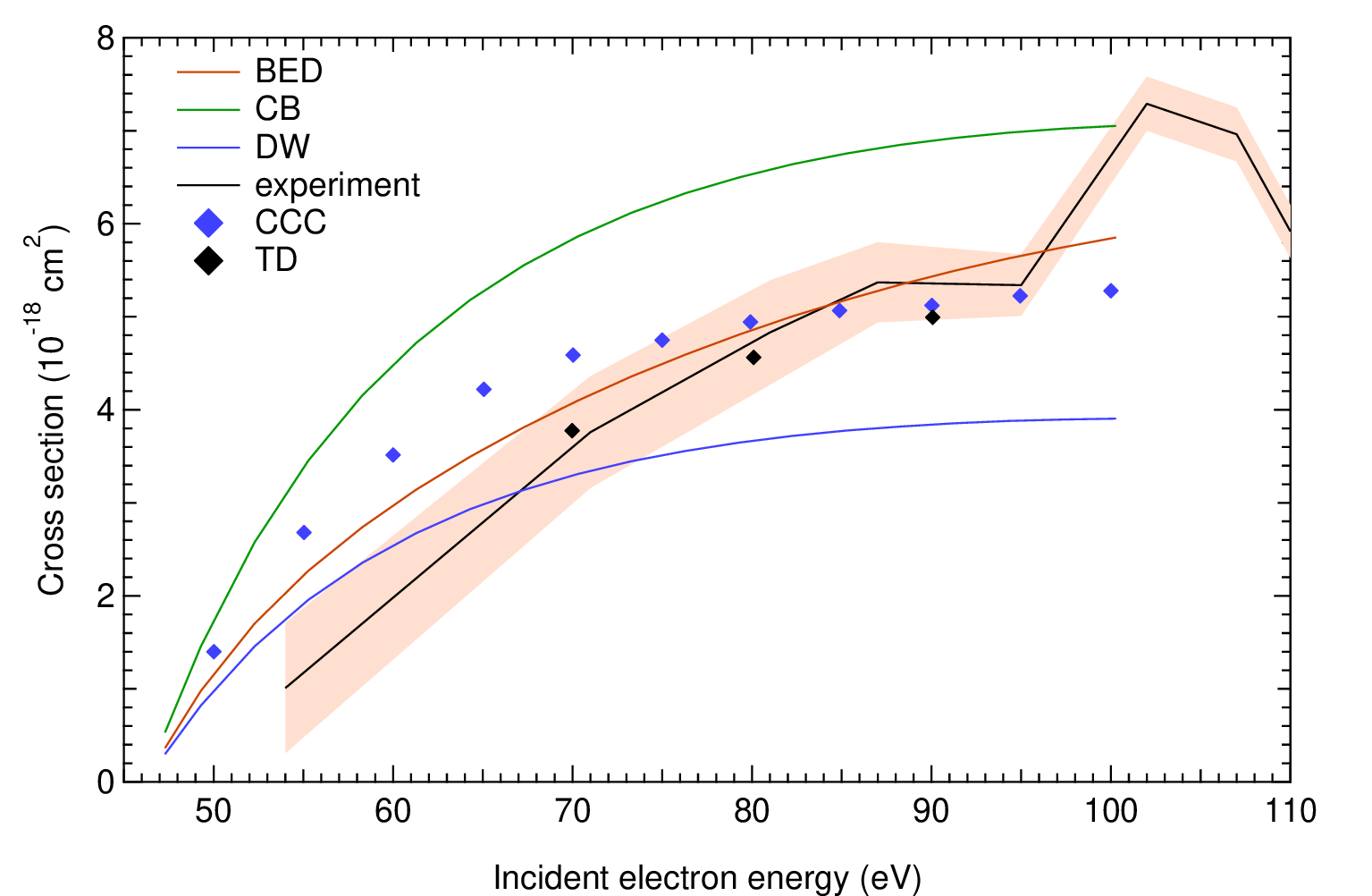}
\caption{Si$^{3+}$ cross sections comparisons. Our results (using the FAC code): BED, CB, DW. Experiment of Crandall \emph{et al.} \cite{Crandall1982}, with the shaded area representing the experimental uncertainty. Calculations of Badnell \emph{et al.} \cite{Badnell1998}: CCC (convergent close-coupling), TD (hybrid time-dependent).}\label{Si3+_igor}
\end{figure}

The pseudo R-matrix results of Badnell \emph{et al.} can be seen in Fig. 1 of Ref. \cite{Badnell1998}. They are in good agreement with Crandall \emph{et al.}'s results for incident electron energies up to 20 eV, and then lie much above (and follow the hybrid time-dependent ones). It is worth mentioning that Badnell \emph{et al.} reported a better agreement with the crossed-beams measurements of Peart \emph{et al.} \cite{Peart1991} for Mg$^+$, but the theoretical results still laid 10 \% higher at 50 eV. Some disagreement was noted between the convergent close-coupling (CCC), the R-matrix and the time-dependent results. We can see that the BED calculation lies between the CCC and time-dependent Badnell ones.

\begin{figure}[!ht]
\centering
\includegraphics[scale=.45]{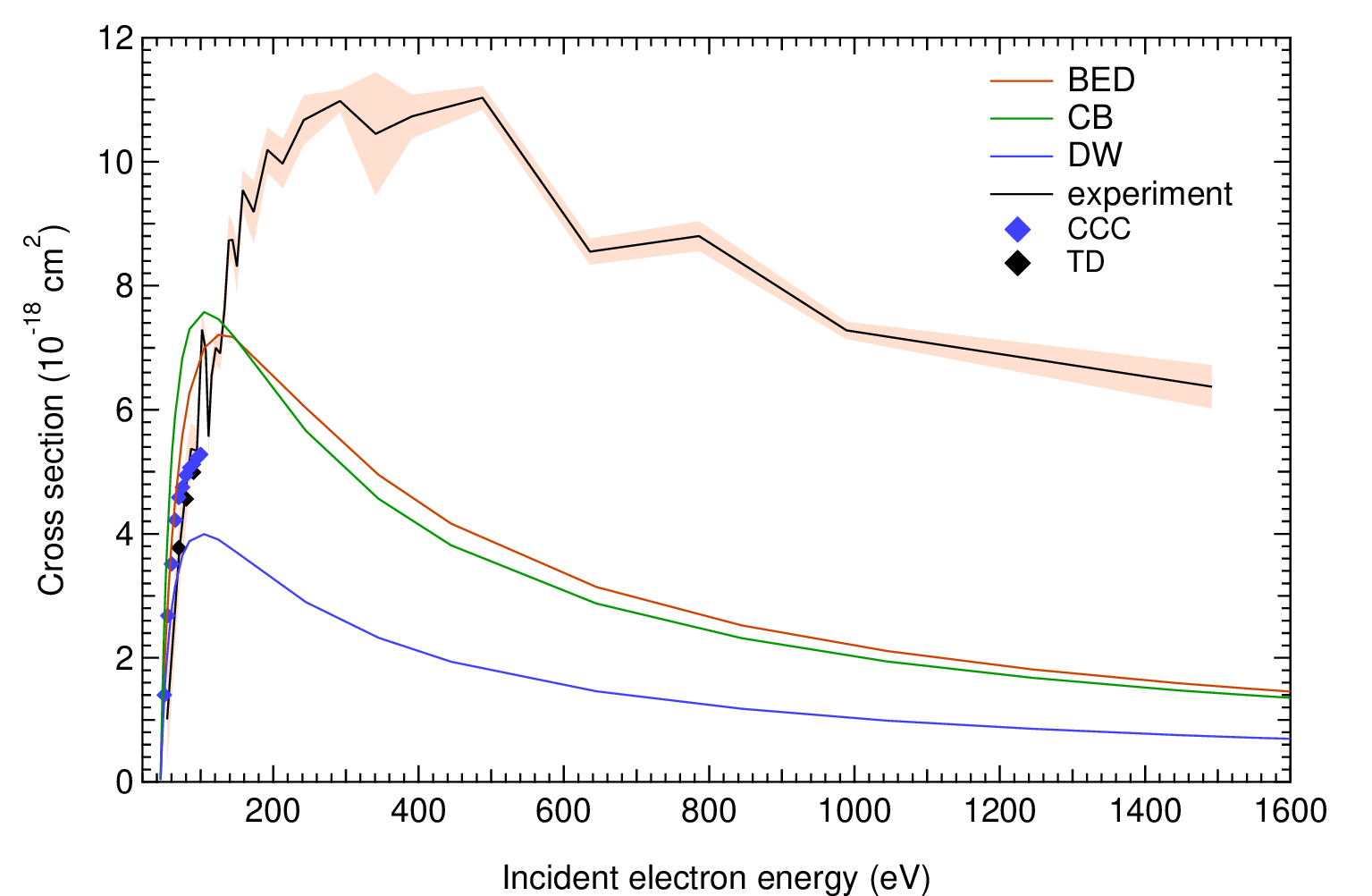}
\caption{Si$^{3+}$ cross section comparisons. Our results with limited CI (using the FAC code): BED, CB, DW. Experiment of Crandall \emph{et al.} \cite{Crandall1982}, with the shaded area representing the experimental uncertainty. Calculations of Badnell \emph{et al.} \cite{Badnell1998}: CCC (convergent close-coupling), TD (hybrid time-dependent).}\label{Si3+_long_noCI}
\end{figure}

In Figures \ref{Si3+_igor} (small energy range, beginning at the ionization threshold) and \ref{Si3+_long_noCI} (wide energy range), we compare our calculations considering only the channel $(1)^2(2)^8(3)^1\rightarrow (1)^2(2)^8$ (the integer value inside the brackets is the principal quantum number, e.g., $(2)$ means the L shell $n=2$, \emph{i.e.}, 2s and 2p), to experimental data of Crandall \emph{et al.} \cite{Crandall1982}. At low energy, CB is too high, and DW too low. At high energy, CB drops, and the slope of the BED cross section seems to be the closest to the experimental one. Our BED results show a better agreement with experiment than the convergent close-coupling (CCC) method (see Fig. \ref{Si3+_igor}).

In order to investigate the effect of the configuration interaction (CI) on cross sections, we consider an additional channel, resulting in the set: 

\begin{equation*}
    (1)^2(2)^8(3)^1\rightarrow \left\{\begin{array}{l}
    (1)^2(2)^8\\
    (1)^2(2)^7(3)^1\\
    \end{array}\right.
\end{equation*}

\begin{figure}[!ht]
\centering
\includegraphics[scale=.45]{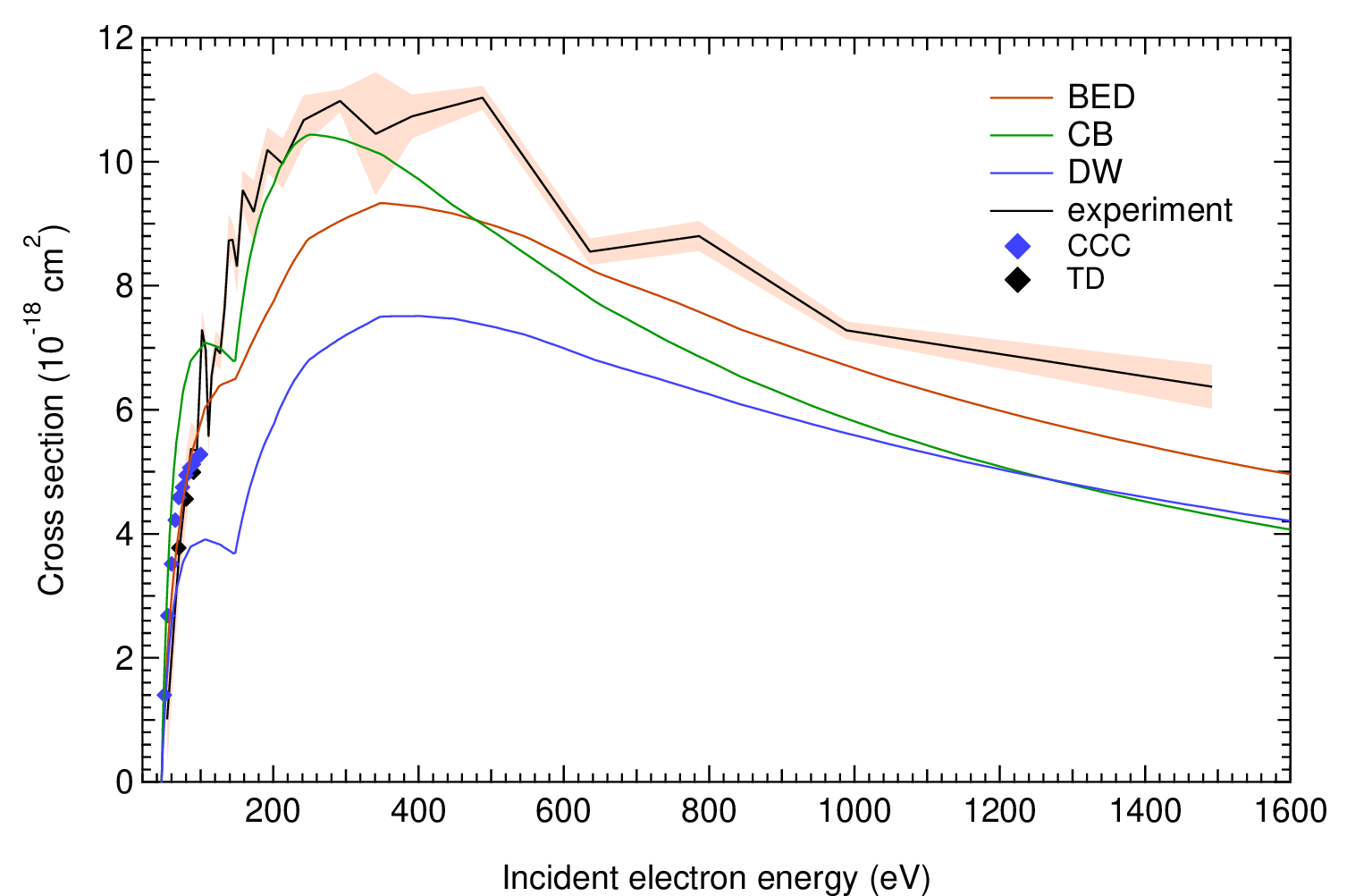}
\caption{Si$^{3+}$ cross section comparisons. Our results with extended CI (using the FAC code): BED, CB, DW. Experiment of Crandall \emph{et al.} \cite{Crandall1982}, with the shaded area representing the experimental uncertainty. Calculations of Badnell \emph{et al.} \cite{Badnell1998}: CCC (convergent close-coupling), TD (hybrid time-dependent).}\label{Si3+_long_CI}
\end{figure}

\begin{figure}[!ht]
\centering
\includegraphics[scale=.45]{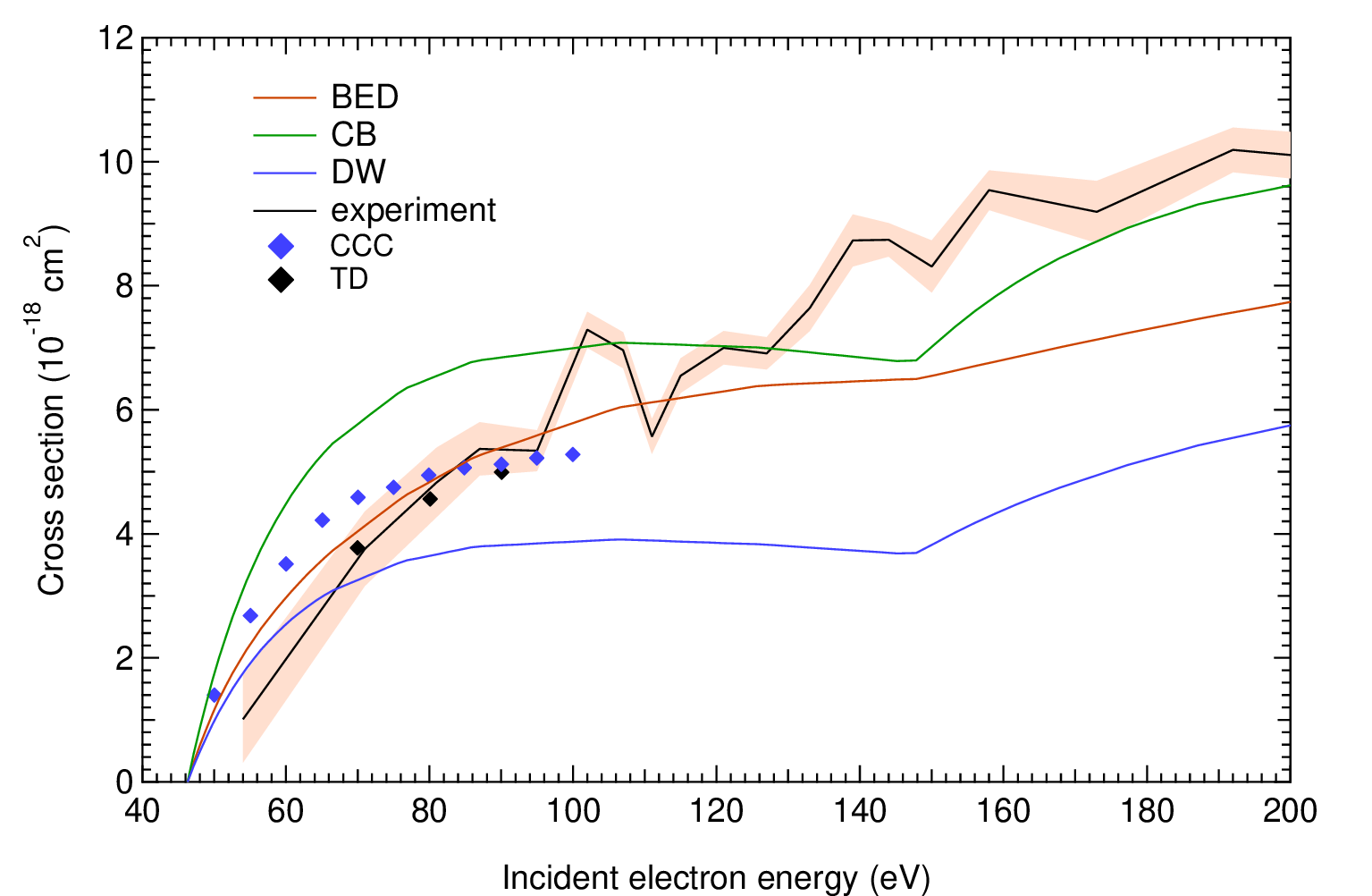}
\caption{Si$^{3+}$ cross section comparisons. Zoom in Fig. \ref{Si3+_long_CI} for the lowest impact energies. Our results with extended CI (using the FAC code): BED, CB, DW. Experiment of Crandall \emph{et al.} \cite{Crandall1982}, with the shaded area representing the experimental uncertainty. Calculations of Badnell \emph{et al.} \cite{Badnell1998}: CCC (convergent close-coupling), TD (hybrid time-dependent).}\label{Si3+_long_CI_zoom}
\end{figure}

It is worth emphasizing that since the FAC calculations are carried out on the basis of Layzer complexes (shells), the most general configuration interaction is included. This means that configuration mixing is not limited to relativistic subconfigurations ($n\ell j$) of a non-relativistic subshell ($n\ell$), but applies to all relativistic configurations included in a list of Layzer complexes (representing configurations of a $n$ shell with fixed parity). Thus, several ionization channels are \emph{de facto} taken into account. As an example, in $(1)^2(2)^8(3)^1\rightarrow (1)^2(2)^8$, the electron in the L-shell can be ejected from $(3s)$, $(3p)$ or $(3d)$ subshells. Moreover, as $(2)^7$ is a super-configuration representing 2s$^2$2p$^5$ and 2s$^1$2p$^6$, $(1)^2(2)^8(3)^1$ contains the configurations 2s$^1$2p$^6$3s$^1$, 2s$^2$2p$^5$3s$^1$, 2s$^1$2p$^6$3p$^1$, 2s$^2$2p$^5$3p$^1$, 2s$^1$2p$^6$3d$^1$ and 2s$^2$2p$^5$3d$^1$.

When we account for more final states (and thus extended configuration interaction), the agreement with experiment is improved (see Figs. \ref{Si3+_long_CI} and \ref{Si3+_long_CI_zoom}). Coulomb-Born-Exchange is better at low energy, up to the maximum of the cross section, but BED provides the best overall global agreement. Distorted-Wave is not satisfactory, although it exhibits a right slope at high energy. If BED was slightly shifted to higher energies, it would be in the experimental error bars at low energy and would have a better asymptotic behavior at high energy. The small shoulder at low energy comes from the superposition of different cross sections, namely the ones from K$^2$L$^8$ 3s$^1$ ($J=1/2$) to K$^2$L$^8$ for the lowest ionization energy (around 46 eV) and from K$^2$L$^8$ 3s$^1$ ($J=1/2$) to levels ``originating'' from $(1)^2(2)^7(3)^1$ (in the sense of the purity of an atomic state obtained by taking the trace of the square of the reduced density operator) for the second ionization energy (around 147 eV). The first (and dominant) ones are indicated in Table \ref{tab:lev}. The bumps are also clearly visible in the experiment.

\begin{table}
\centering
\begin{tabular}{ccc}\hline\hline
Energy & $J$ & Relativistic configuration \\ \hline\hline
147.172 & 2 & 2p$_{1/2}^2$ 2p$_{3/2}^3$ 3s$^1$\\
147.495 & 1 & 2p$_{1/2}^2$ 2p$_{3/2}^3$ 3s$^1$\\
147.788 & 0 & 2p$_{1/2}^1$ 2p$_{3/2}^4$ 3s$^1$\\ 
148.542 & 1 & 2p$_{1/2}^1$ 2p$_{3/2}^4$ 3s$^1$ \\\hline\hline
\end{tabular}
\caption{Levels from configuration 1s$^2$2s$^2$2p$^5$3s$^1$ belonging to the super-configuration $(1)^2(2)^7(3)^1$. The energies (in eV) are relative to the energy of the ground level K$^2$L$^8$ 3s$^1$ ($J=1/2$).}\label{tab:lev}
\end{table}

\begin{figure}[!ht]
\centering
\includegraphics[scale=.45]{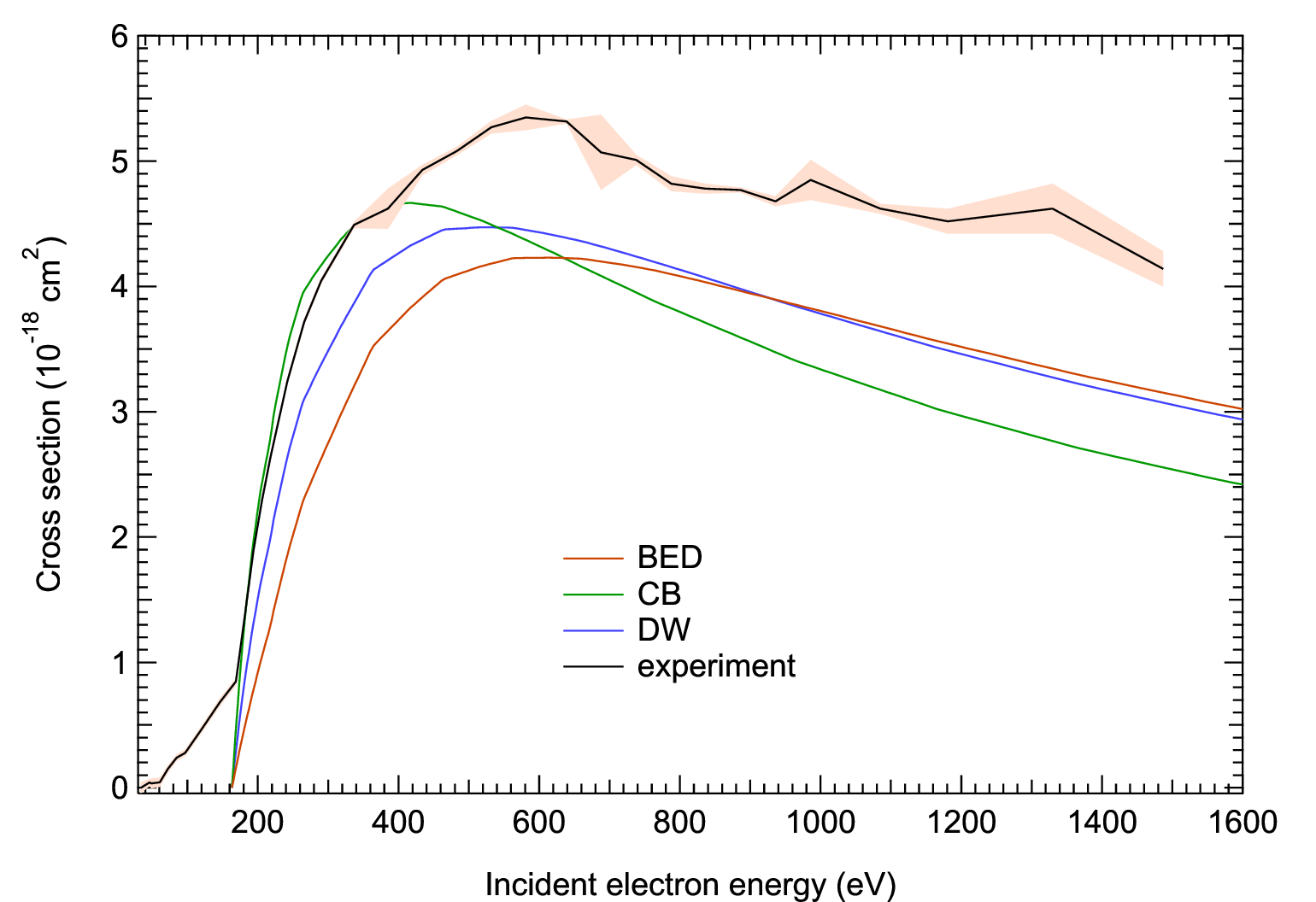}
\caption{Si$^{4+}$ cross section comparisons. Our results with extended CI (using the FAC code): BED, CB, DW. Experimental data are from Thompson \emph{et al.} \cite{Thompson1994}. The shaded area represents the experimental uncertainty.}\label{Si4+_long_CI_igor}
\end{figure}

In the case of Si$^{4+}$, we compare our calculations with the experimental data of Thompson \emph{et al.} \cite{Thompson1994}. We consider the ionization channels:

\begin{equation}
    (1)^2(2)^8\rightarrow \left\{\begin{array}{l}
    (1)^2(2)^7\\
    (1)^2(2)^6(3)^1\\
    \end{array}\right.
\end{equation}
We observe in Fig. \ref{Si4+_long_CI_igor} the same trend as for Si$^{3+}$, except that the effect of extended CI among the final accessible states is very small. If we consider only the ionization channel $(1)^2(2)^8\rightarrow (1)^2(2)^7$, we obtain similar results, due to the fact that the final super-configuration $(1)^2(2)^7$ contains both 2s$^1$2p$^6$ and 2s$^2$2p$^5$, which bring the dominant contributions to the cross section. Thus, although the final super-configuration contains more configurations than in the Si$^{3+}$ case (2s$^0$2p$^6$3s$^1$, 2s$^0$2p$^6$3p$^1$, 2s$^0$2p$^6$3d$^1$, 2s$^1$2p$^6$, 2s$^1$2p$^5$3s$^1$, 2s$^1$2p$^5$3p$^1$, 2s$^1$2p$^5$3d$^1$, 2s$^2$2p$^5$, 2s$^2$2p$^4$3s$^1$, 2s$^2$2p$^4$3p$^1$ and 2s$^2$2p$^4$3d$^1$), adding $(1)^2(2)^6(3)^1$ has a small effect on the results. All approaches underestimate the experimental cross section. The CB method seems well suited for low incident electron energies, but the asymptotic behaviour is better reproduced by the BED method. DW is likely to yield the best compromise. 

\begin{figure}[!ht]
\centering
\includegraphics[scale=.45]{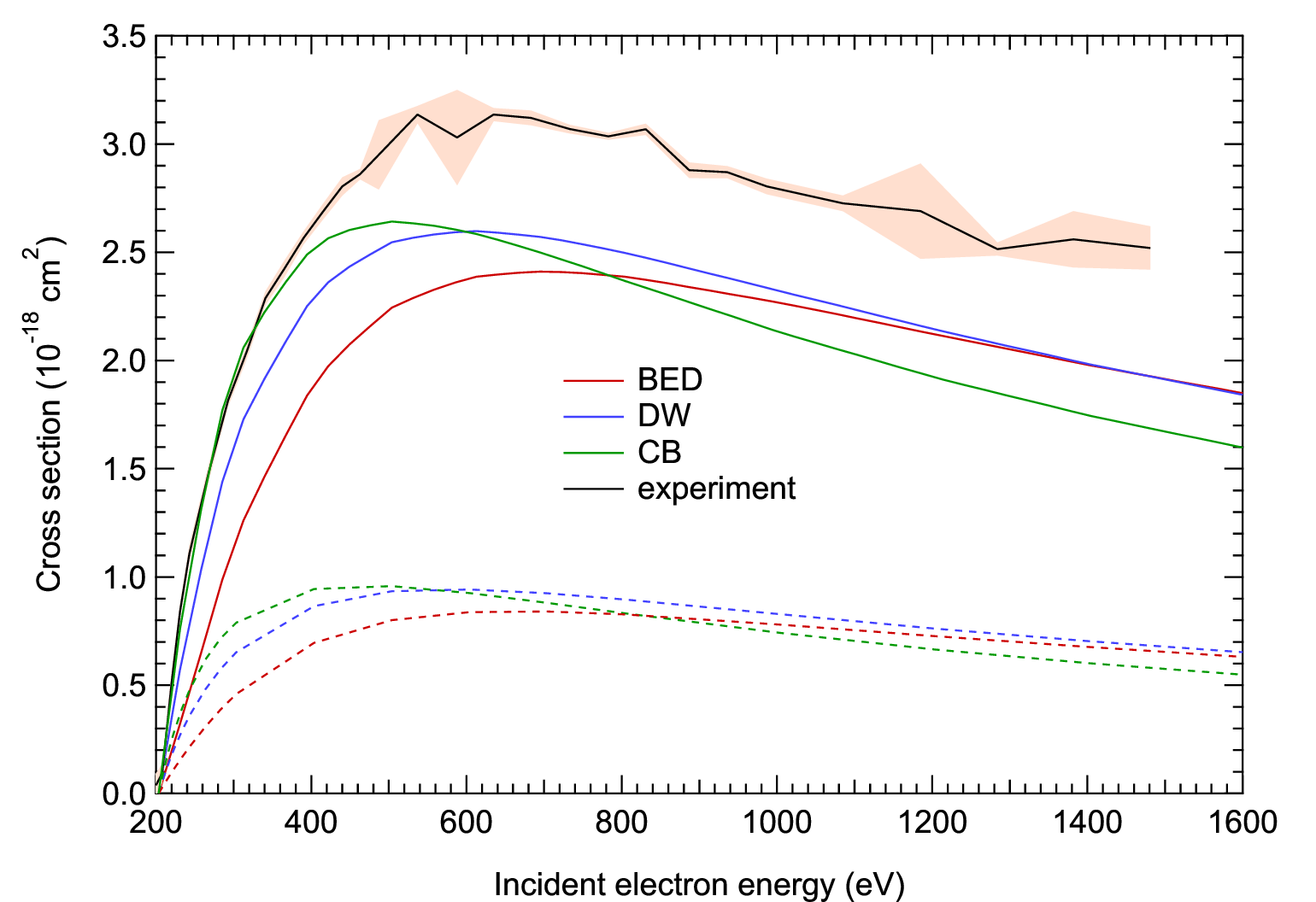}
\caption{Si$^{5+}$ cross section comparisons. Our results with limited CI (using the FAC code): BED, CB, DW. Experimental data are from Thompson \emph{et al.} \cite{Thompson1994}. The shaded area represents the experimental uncertainty}.\label{Si5+_long_noCI_igor}
\end{figure}

As in the previous case, for Si$^{5+}$ ions experimental data are from Thompson \emph{et al.} \cite{Thompson1994}. Let us consider the ionization channel $(1)^2(2)^7\rightarrow (1)^2(2)^6$. As shown in Fig. \ref{Si5+_long_noCI_igor}, the DW method provides the best agreement with experiment.

\begin{figure}[!ht]
\centering
\includegraphics[scale=.45]{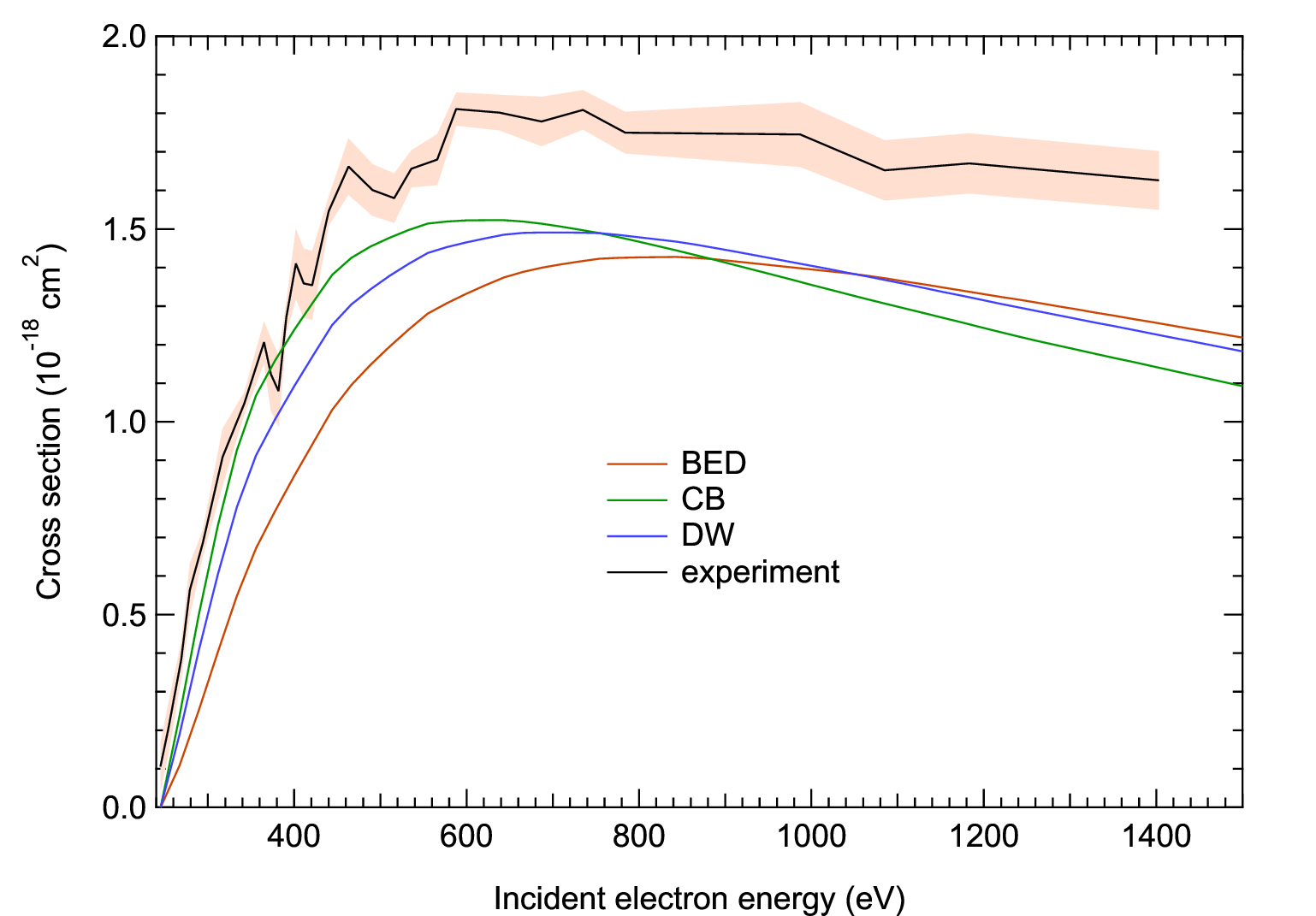}
\caption{Si$^{6+}$ cross section comparisons. Our results with limited CI (using the FAC code): BED, CB, DW. Experimental data are from Zeijlmans van Emmichoven \emph{et al.} \cite{Zeijlmans1993}. The shaded area represents the experimental uncertainty}.\label{Si6+_long_noCI_igor}
\end{figure}

For Si$^{6+}$, we compare our calculations to the experimental data of Zeijlmans van Emmichoven \emph{et al.} \cite{Zeijlmans1993}. We calculate the cross section associated with the super transition array $(1)^2(2)^6\rightarrow (1)^2(2)^5$. The results of the analysis and the conclusions are roughly the same as for Si$^{5+}$ (see Fig. \ref{Si6+_long_noCI_igor}).

\begin{figure}[!ht]
\centering
\includegraphics[scale=.45]{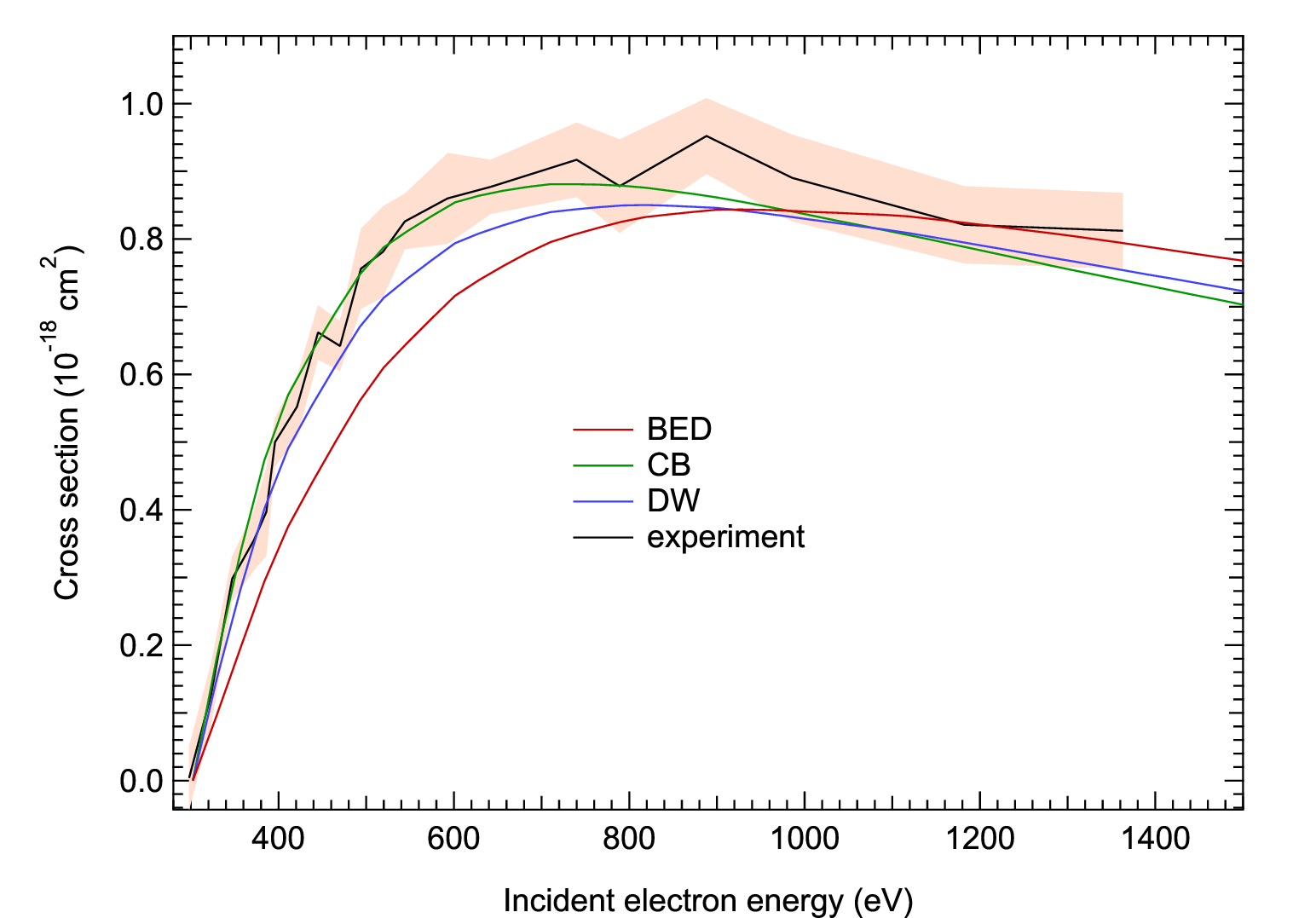}
\caption{Si$^{7+}$ cross section comparisons. Our results with limited CI (using the FAC code): BED, CB, DW. Experimental data are from Zeijlmans van Emmichoven \emph{et al.} \cite{Zeijlmans1993}. The shaded area represents the experimental uncertainty}.\label{Si7+_long_noCI_igor}
\end{figure}

For the last ion Si$^{7+}$, as in the previous case, we compare our calculations to the measurements of Zeijlmans van Emmichoven \emph{et al.} \cite{Zeijlmans1993}. We consider the transition array $(1)^2(2)^5\rightarrow (1)^2(2)^4$. As can be seen in Fig. \ref{Si7+_long_noCI_igor}, the BED results are too low at low energy, but very satisfactory at high energy. At low energy, CB fits the upper bounds of the experimental error bars, and DW the lowest. DW follows the lower bounds of the error bars very closely. Up to the maximum, the CB cross section is in better agreement with experiment than DW and BED, but at intermediate to high energies, DW becomes closer to experimental values than CB or BED. The DW method seems to be the one that shows the best overall agreement. 

In the following, we calculate the rate coefficient of silicon ions over a large temperature interval and compare our results to experiments.

\section{Rate coefficient calculations}\label{sec3}

\subsection{Analytical representation of measured and calculated cross sections}\label{subsec31}

Different analytical formulas were proposed in order to represent the electron-impact ionization cross section. Typically, the search for such formulas is driven by the goal of obtaining values that closely match reference (measured or computed) data, while simplifying the integration of the cross section over the Maxwell-Boltzmann distribution of free electrons. This simplification is crucial for obtaining rate coefficients, which are key components in the collisional-radiative modeling of NLTE plasmas. For instance, the formulations of Lotz \cite{Lotz1968} and Younger \cite{Younger1982} have been widely used in the past. Two decades ago, Bernshtam \textit{et al.} \cite{Bernshtam2000} published an empirical formula, inspired from the Lotz one, involving two parameters depending on the orbital quantum number of the initial state and adjusted to fit experimental results. In a recent work, we proposed a semi-empirical formula for the cross section, depending on four adjustable parameters determined by fitting measured carbon, nitrogen and oxygen ions and neutral aluminum cross sections. It reads \cite{Benredjem2024}:
\begin{equation}\label{fit2}
    \sigma(E)=A\,\frac{\ln(E/E_i)}{E/E_i}\,\sum_{p=0}^N\frac{B_p}{(E/E_i)^p},
\end{equation}
with $B_0=1$. $\sigma$ fulfills the condition $\sigma(E_i)=0$, as it should. The knowledge of the adjustable parameters enables one to derive analytic expressions of the rate coefficients and to calculate cross sections in energy ranges where no experimental results are available. In the following, we fit experimental or calculated cross sections to $\sigma$.

We focus on the highly-ionized atoms Si$^{5+}$, Si$^{6+}$ and Si$^{7+}$ since at temperatures occurring in ICF plasmas, the populations corresponding to lower ion charges are negligible. Also, it is worth stressing that the FAC code is more appropriate for highly-ionized atoms than for quasi-neutral ones. The parameter values obtained by fitting experimental or calculated (BED, CB, DW) to $\sigma(E)$ are given in Tables \ref{table_rate_exp}, \ref{table_rate_bed}, \ref{table_rate_cb} and \ref{table_rate_dw}, respectively.

\begin{table}[!ht]
\begin{center}
\begin{tabular}{crrr}\hline
fitting parameter & Si$^{5+}$ & Si$^{6+}$ & Si$^{7+}$\\\hline\hline
$A$ & 9.7090 & 5.9389& 1.9263\\
$B_0$ & 1 & 1 & 1\\
$B_1$ & -0.414451 & -0.663237 & 1.316880\\ 
$B_2$ & -0.003916 & 0.418125 & -1.488860\\
$B_3$ & -0.000030 & -0.000011 & 0.000388\\\hline
\end{tabular}
\end{center}
\caption{Values of the adjustable parameters in equation (\ref{fit2}), obtained by fitting measured cross sections. $A$ is in 10$^{-18}$ cm$^2$, $B_0, ..., B_3$ are dimensionless.}\label{table_rate_exp}
\end{table}

\begin{table}[!ht]
\begin{center}
\begin{tabular}{rrrr}\hline
Fitting parameter & Si$^{5+}$ & Si$^{6+}$ & Si$^{7+}$\\\hline\hline
$A$ & 7.8009 & 4.5605 & 2.5804\\
$B_0$ & 1 & 1 & 1\\
$B_1$ & -0.354498 & -0.274663 & -0.167129\\ 
$B_2$ & -0.453230 & -0.517641 & -0.446714\\
$B_3$ & -0.000090 & -0.000177 & -0.000201\\\hline
\end{tabular}
\end{center}
\caption{Values of the adjustable parameters in equation (\ref{fit2}), obtained by fitting BED cross sections. $A$ is in 10$^{-18}$ cm$^2$, $B_0, ..., B_3$ are dimensionless.}\label{table_rate_bed}
\end{table}

\begin{table}[!ht]
\begin{center}
\begin{tabular}{rrrr}\hline
Fitting parameter & Si$^{5+}$ & Si$^{6+}$ & Si$^{7+}$\\\hline\hline
$A$ & 5.6619 & 3.1685 & 1.8919\\
$B_0$ & 1 & 1 & 1\\
$B_1$ & 1.051860 & 1.320620 & 0.987895\\ 
$B_2$ & -0.936187 & -1.326620 & -0.763929\\
$B_3$ & -0.000087 & -0.000057 & -0.000133\\\hline
\end{tabular}
\end{center}
\caption{Values of the adjustable parameters in equation (\ref{fit2}), obtained by fitting CB cross sections. $A$ is in 10$^{-18}$ cm$^2$, $B_0, ..., B_3$ are dimensionless.}\label{table_rate_cb}
\end{table}

\begin{table}[!ht]
\begin{center}
\begin{tabular}{rrrr}\hline
Fitting parameter & Si$^{5+}$ & Si$^{6+}$ & Si$^{7+}$\\\hline\hline
$A$ & 7.2607 & 3.9123 & 2.1388 \\
$B_0$ & 1 & 1 & 1\\
$B_1$ & 0.090841 & 0.394576 & 0.439330\\ 
$B_2$ & -0.502968 & -0.803185 & -0.592295\\
$B_3$ & -0.000074 & -0.000093 & -0.000133\\\hline
\end{tabular}
\end{center}
\caption{Values of the adjustable parameters in equation (\ref{fit2}), obtained by fitting DW cross sections. $A$ is in 10$^{-18}$ cm$^2$, $B_0, ..., B_3$ are dimensionless.}\label{table_rate_dw}
\end{table}

\subsection{Rate calculations}\label{subsec32}

In the following, we concentrate on the EII rate coefficient, which is expressed as
\begin{equation}
    q=\int_{E_i}^{\infty} \sigma(E)\,v\,\rho(E)\,dE,
\end{equation}
where $v=\sqrt{2E/m_e}$ is the velocity of the incident electron, with $m_e$ the electron mass, $E_i$ the ionization energy and $\rho$ the normalized distribution of the free electrons:
\begin{equation*}
    \rho(E)=\frac{2}{\sqrt{\pi}}\frac{1}{(k_{\rm B}T_e)^{3/2}}\sqrt{E}~e^{-E/(k_{\rm B}T_e)},\label{B-distribution}
\end{equation*} 
where $T_e$ is the electron temperature. In the case where the Boltzmann statistics is relevant, the rate coefficient becomes
\begin{equation}
    q=\sqrt{\frac{8\,k_{\rm B}T_e}{\pi\,m_e}}\,\int_b^{\infty}t\,\sigma(t)\, e^{-t}\,dt,
\end{equation}
where $t=E/(k_{\rm B}T_e)$ and $b=E_i/(k_{\rm B}T_e)$. The integral can be calculated analytically giving the rate coefficient. We have:
\begin{eqnarray*}
    q&=&\sqrt{\frac{8\,k_{\rm B}T_e}{{\pi\,m_e}}}\,\int_b^{\infty}t\,\sigma(t)\, e^{-t}\,dt\nonumber\\
    &=&\sqrt{\frac{8\,k_{\rm B}T_e}{{\pi~m_e}}}\,b~A\sum_{p=0}^3B_p~\int_b^{\infty}\ln\left(\frac{t}{b}\right)\frac{1}{(t/b)^{p}}\,e^{-t}\,dt\nonumber\\
    &=&\sqrt{\frac{8\,k_{\rm B}T_e}{{\pi~m_e}}}\,b^2\,A\sum_{p=0}^3B_p~\int_1^{\infty}~\ln(x)\frac{1}{x^p}\,e^{-b\,x}\,dx.
\end{eqnarray*}
The rate coefficient can then be expressed in terms of the generalized integro-exponential functions \cite{Milgram1985,MacLeod2002,Luke1969}:
\begin{equation*}
    E_p^j(b)=\frac{1}{\Gamma(j+1)}\int_1^{\infty} [\ln(x)]^j\,\frac{1}{x^p}e^{-b\,x}\,dx,
\end{equation*}
giving
\begin{equation}
    q=\sqrt{\frac{8\,k_{\rm B}T_e}{\pi\,m_e}}\,b^2\,A\sum_{p=0}^3B_p\,\Gamma(p+1)E_p^1(b),
\end{equation}
where $\Gamma$ is the Gamma function. The functions $E_p^1$ are easily calculated. We have in particular
\begin{eqnarray*}
    E_0^1(b)&=&\int_1^{\infty}\ln(x)\,e^{-bx}\,dx=\frac{1}{b}\,\Gamma(0,b)
\end{eqnarray*}
and
\begin{equation}
    E_1^1(b)=\int_1^{\infty}\frac{\ln(x)}{x}\,e^{-bx}\,dx=G_{2,3}^{3,0}\left(b\left|\begin{array}{c}
    1,1\\
    0, 0,0
    \end{array}\right.\right),
\end{equation}
where $G$ is a Meijer $G$-function \cite{Prudnikov1990}.
The other integrals involved in the rate coefficients are easily obtained by using the recurrence relation \cite{Milgram1985}:
\begin{equation}
    E_p^1(b)=\frac{b\,E_{p-1}^1(b)-E_{p}^0(b)}{1-p}=\frac{b\,E_{p-1}^1(b)-E_{p}(b)}{1-p}\hspace{2cm} p\neq 1.
\end{equation}

\begin{figure}[!ht]
\centering
\includegraphics[scale=.45]{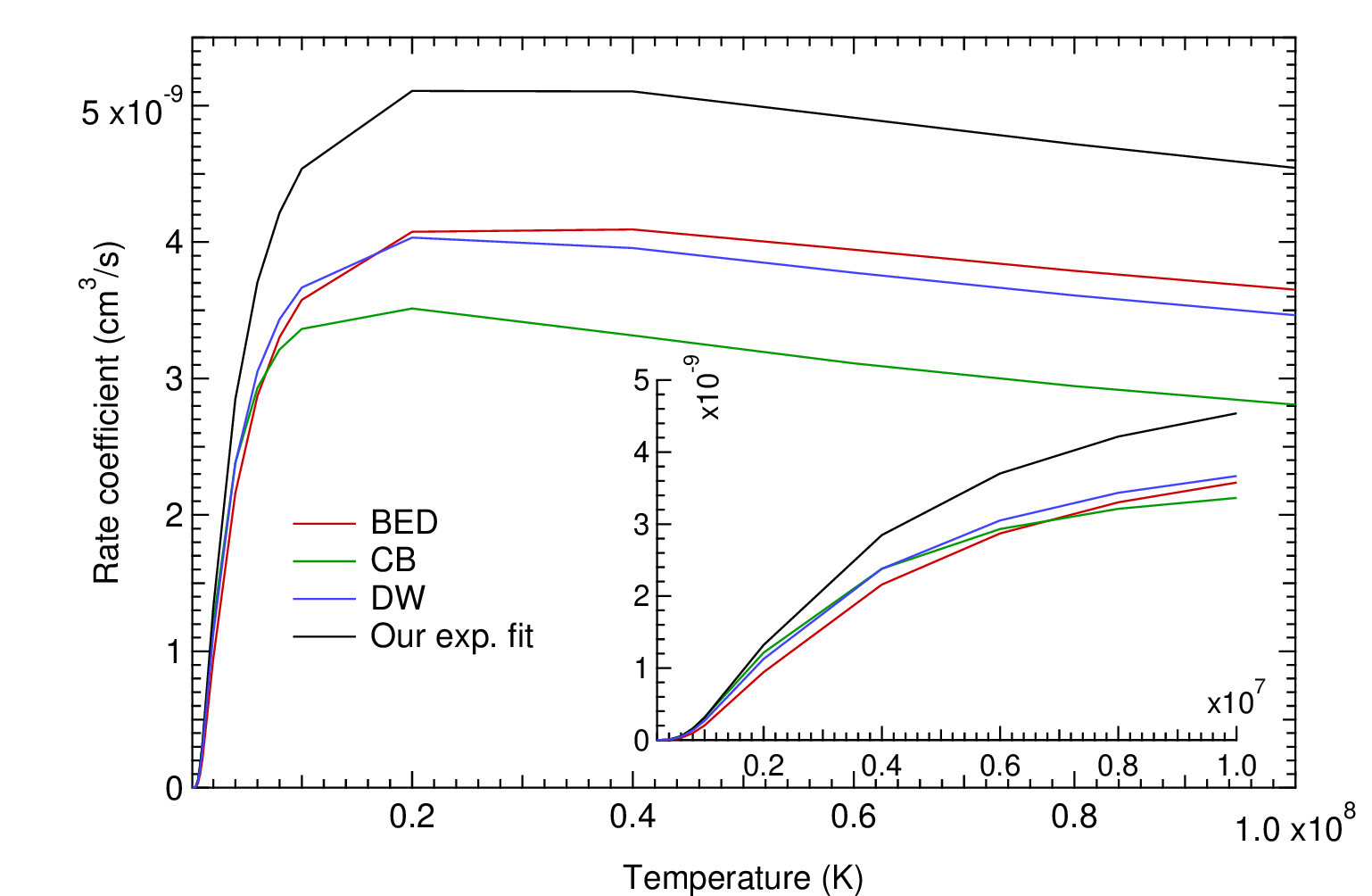}
\caption{Rate coefficient of Si$^{5+}$ as a function of the temperature. The solid lines represent our fits of the experiment (Our exp. fit) and of the BED, CB and DW calculations}.\label{Taux_Si5+}
\end{figure}

\begin{figure}[!ht]
\centering
\includegraphics[scale=.45]{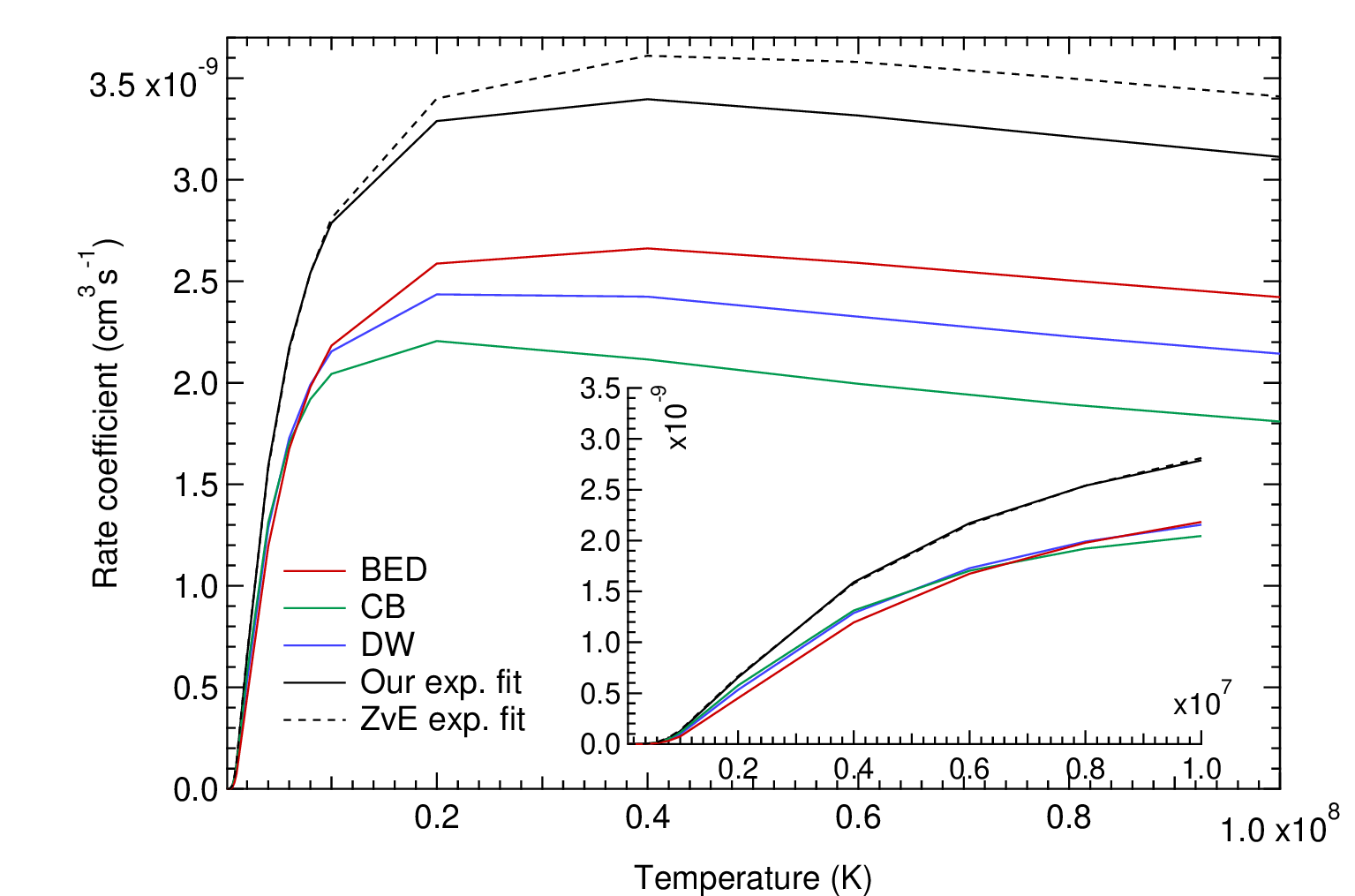}
\caption{Rate coefficient of Si$^{6+}$ as a function of the temperature. The solid lines represent our fits of the experiment (Our exp. fit) and of the BED, CB and DW calculations. The dashed line represents the fit of the experiments made by Zeijlmans van Emmichoven \emph{et al.} (ZvE exp. fit).}\label{Taux_Si6+}
\end{figure}

\begin{figure}[!ht]
\centering
\includegraphics[scale=.45]{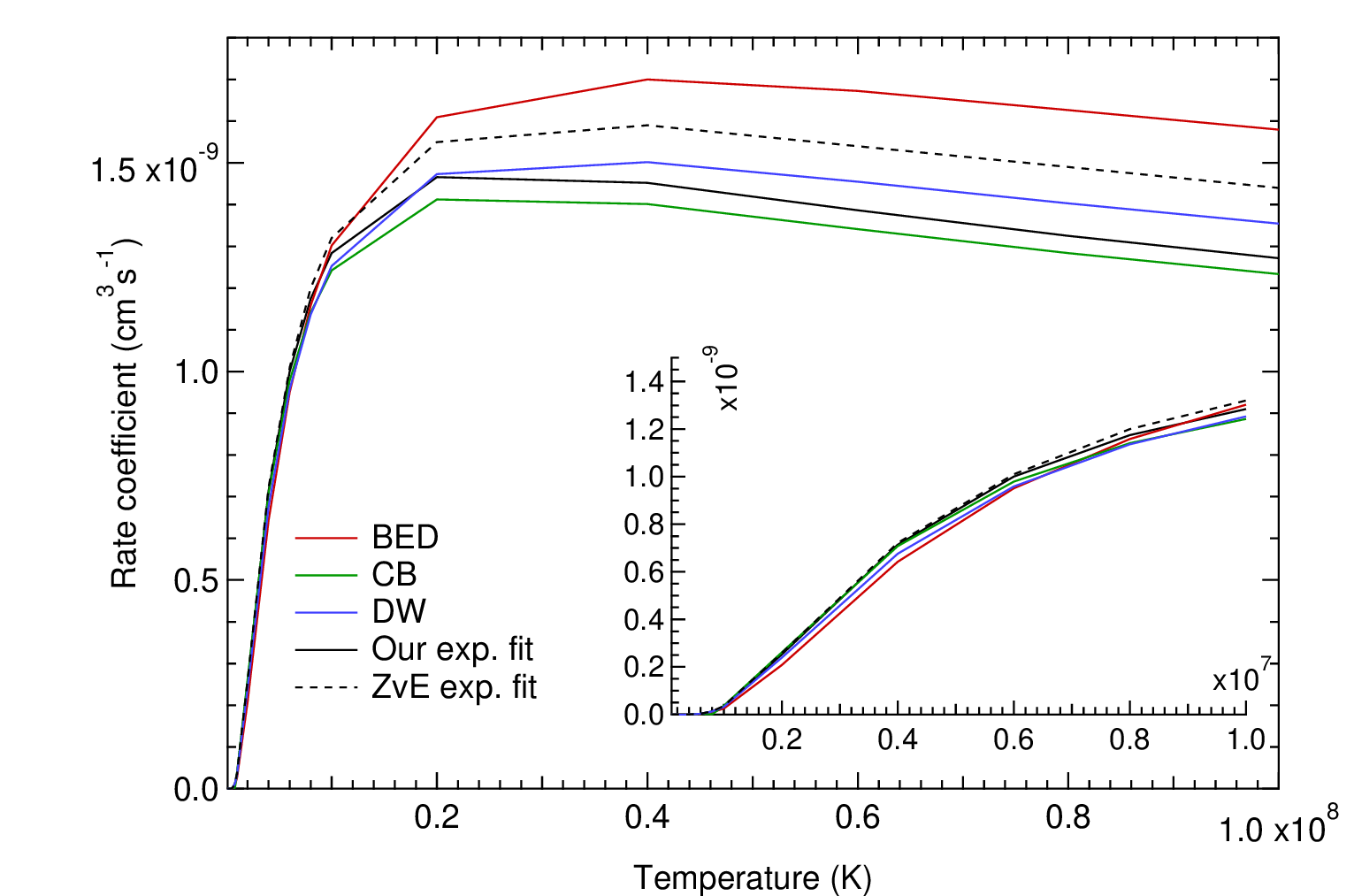}
\caption{Rate coefficient of Si$^{7+}$ as a function of the temperature. The solid lines represent our fits of the experiment (Our exp. fit) and of the BED, CB and DW calculations. The dashed line represents the fit of the experiments made by Zeijlmans van Emmichoven \emph{et al.} (ZvE exp. fit).}\label{Taux_Si7+}
\end{figure}

Figure \ref{Taux_Si5+} represents the variation with temperature of the rate coefficient for Si$^{5+}$. Plasmas of interest in fusion studies are within the temperature range. We can see that the CB method provides the best agreement with experiment at low temperatures (below 3$\times10^6$ K). The DW method is the most appropriate for temperatures from 3$\times10^6$ K to 20 MK. The BED method is the closest to measurements above 20 MK. The behaviour is similar for Si$^{6+}$ (see Figure \ref{Taux_Si6+}). Concerning Si$^{7+}$ (Figure \ref{Taux_Si7+}), we can see that the three cross sections are closer to measurements than Si$^{5+}$ and Si$^{6+}$. The DW method presents the best agreement at low temperatures and above 20 MK while the CB method is closer to experimental results when 2 $< T_e <$ 7 MK.

In the following, we compare our calculations with the work of Zeijlmans van Emmichoven \emph{et al.} \cite{Zeijlmans1993}. These authors investigated Maxwellian rate coefficients for Si$^{6+}$ and Si$^{7+}$. They proposed an analytical approximation, enabling one to compute the rate either from an expansion in Chebyshev polynomials, or using the Clenshaw algorithm.

\subsection{The approach of Zeijlmans van Emmichoven \emph{et al.}}\label{subsec33}

In contrast to our work, Zeijlmans van Emmichoven \emph{et al.} proposed a fitting formula for rate coefficients rather than cross sections. When using Chebyshev polynomials $T_j(x)$ \cite{Rivlin1974}, their rate coefficient reads:

\begin{table}
\begin{center}
\begin{tabular}{ccc}\hline
Fitting parameter & Si$^{6+}$ & Si$^{7+}$\\\hline\hline
$a_0$ & 1648.96 & 641.257\\
$a_1$ & -127.945 & 75.0597\\ 
$a_2$ & -419.445 & -333.790\\
$a_3$ & -50.4497 & 23.1258\\
$a_4$ & 188.404 & 105.003\\
$a_5$ & -69.4733 &-38.7064\\
$a_6$ & -7.75955 & -13.4105\\
$a_7$ & 19.0429 & 13.9309\\
$a_8$ & -7.71838 & -2.57181\\
$a_9$ & 0.00000 & -1.69028\\
$a_{10}$ & 0.00000 & 0.933665\\\hline
\end{tabular}
\end{center}
\caption{Rate coefficient fitting parameters proposed by Zeijlmans van Emmichoven \emph{et al.} \cite{Zeijlmans1993}. All parameters are in units of 10$^{-15}$ cm$^3\cdot$ K$^{-1/2}\cdot$ s$^{-1}$. Rate coefficients in the range $10^5\leq T_e\leq 10^8$ K may be computed using these parameters directly in a Chebyshev polynomial expansion \cite{Rivlin1974}, or through the Clenshaw algorithm \cite{Clenshaw1955,Clenshaw1957}.}\label{table_rate}
\end{table}

\begin{equation}
    q_{\rm ZvE}(T_e)=T_e^{1/2}e^{-b}\sum_{j=0}^na_jT_j(x),
\end{equation}
where $b$ is the reduced ionization potential defined in \ref{subsec32} and $x=(\log_{10}T_e-6.5)/1.5$, $T_e$ being expressed in K. The $a_j$ values relative to Si$^{6+}$ and Si$^{7+}$ are given in Table \ref{table_rate}. Zeijlmans van Emmichoven \emph{et al.} claim that the $a_j$ values reproduce the rate coefficients to with in 1 \% over the temperature range considered here. 

The Clenshaw algorithm leads to an expression of the rate coefficient as 
\begin{equation}
    q_{\rm C}(T_e)=\frac{\sqrt{T_e}}{2}e^{-b}\left(b_0-b_2\right),
\end{equation}
with the coefficients $b_0$ and $b_2$ calculated using the Clenshaw algorithm \cite{Clenshaw1955,Clenshaw1957}. The latter is presented as a downward recurrence in which the coefficients satisfy $b_n=2x\,b_{n+1}-b_{n+2}+a_n$ with $b_{11}=b_{12}=0$. Here, we present it in an upward way, since it is easier to implement in a computer algebra system or a numerical code. Setting $c_m=b_{12-m}$, we obtain $c_m=2x\,c_{m+1}-c_{m+2}+a_{12-m}$ or $c_{m+2}=2x\,c_{m+1}-c_m+a_{12-m}$ with $c_0=c_1=0$. Note that the Clenshaw algorithm works only if $a_0$ is replaced by $2a_0$ in the expression of $b_0$, \emph{i.e.}, $b_0=2a_0+2x\,b_1-b_2$ which, for the $c_m$ coefficients means $c_{12}=2a_0+2x\,c_{11}-c_{10}$. We believe that the latter point is worth mentioning since it is often eluded by authors. 

As can be seen in Figs. \ref{Taux_Si6+} and \ref{Taux_Si7+}, the fit of the measurements of Zeijlmans van Emmichoven \emph{et al.} (ZvE, black dashed line) yields rates which are systematically higher than ours (black solid line). In the Si$^{6+}$ case, the closest calculation to ZvE is still BED, but in the Si$^{7+}$ case, their rates lie between DW and BED. Our conclusions thus differ from theirs in the latter case. This is due to the fact that the assumptions and methodology underlying their fitting procedure differ significantly from ours, and tends to indicate that the quality and reliability of the fitting formula deserves the greatest attention. 

In Ref. \cite{Benredjem2024}, we compared different cross sections with each other as well as the corresponding rates. We also investigated density effects by considering different energy distributions for the free electrons: Maxwell-Boltzmann versus Fermi-Dirac distribution. In this work, we use the formulation that has proved to be the best in the above mentioned article, and compare it with measurements and other calculations for silicon ions. Our approach is well-suited for a study of the variation of the rate with temperature, which we compare with that of Zeijlamns van Emmichoven et al. (ZvE). In practice, we fit the cross section, and then integrate it over the Maxwellian distribution, which introduces the dependence with respect of the temperature. In the ZvE approach, the fitting procedure is applied directly to the rate. As a consequence, our results show a better agreement with the experiment than ZvE ones. In our study, the error arises from both experimental uncertainty and the accuracy of the fitting formula used for the cross-section. For ZvE, the error includes the experimental uncertainty, the fitting formula for the rate as a function of temperature, and the propagation of errors in the Clenshaw algorithm. Additionally, the dispersion in our results compared to the theoretical methods (CB, BED, DW) provides insight into the uncertainty inherent in the model.

\section{Conclusion}\label{sec4} 

In this work, we calculated electron-impact ionization cross sections of silicon ions by using different methods. The results were compared to each other as well as to other computations and to experimental data when available. We concentrated on cross sections and rate coefficients of silicon ions of interest in plasmas at high temperatures, namely Si$^{3+}$ to Si$^{10+}$. In all investigated cases, comparisons with other calculations and experiments are highlighted. 

We first calculated cross sections and compared them with the experimental results of Crandall \emph{et al.} \cite{Crandall1982} as well as with convergent close-coupling and time-dependant calculations by Badnell \emph{et al.} \cite{Badnell1998}. It appears that the BED results show the best overall agreement with calculations, and a satisfactory agreement with experiment is noticed when an extended set of final states and configuration interaction are taken into account.

We then compared our calculations for Si$^{4+}$ and Si$^{5+}$ with experimental values from Thompson \emph{et al.} \cite{Thompson1994}, and for Si$^{6+}$ and Si$^{7+}$ to Zeijlmans van Emmichoven \emph{et al.} \cite{Zeijlmans1993}. The configuration interaction effect was shown to be important for Si$^{3+}$, but not really for Si$^{4+}$. It is not easy to clearly identify the best method among CB, DW and BED. However, for the silicon ions studied in this work, the following conclusions can be drawn. First, for all the three methods, the reliability of the results (\emph{i.e.}, their agreement with experiment), increases with the ion charge. We notice that in general the CB method is the most accurate close to the ionization threshold (at low incident electron energy), while the BED method provides the best ``slope'' at high energy, and also yields the best overall agreement. Thus, since the cross sections had to be weighted by the Maxwellian distribution in order to obtain the rates, the best method can be selected according to the temperature value. We have calculated electron-impact ionization rates for Si$^{5+}$, Si$^{6+}$ and Si$^{7+}$, using our previously published analytical representation of the cross section \cite{Benredjem2024} and fitting with CB, BED and DW FAC computations. Then, we compared our results with the rates deduced from the experiment of Zeijlmans van Emmichoven \emph{et al.} \cite{Zeijlmans1993}, using a Chebyshev expansion and the Clenshaw algorithm, for Si$^{6+}$ and Si$^{7+}$. 

In a forthcoming work, we will present cross sections and rates associated with transition arrays, \emph{i.e.} the ensemble of transitions between two given configurations (rather than between two levels). In addition to silicon, we will also study germanium, which can be used as a dopant in the ablator of an indirect-drive capsule.


\end{document}